\documentclass{aastex63}

\usepackage[utf8x]{inputenc}
\usepackage{amssymb}
\usepackage{amsmath}
\usepackage{natbib}
\usepackage{hyperref}

\received{\today}
\revised{\today}
\accepted{\today}
\submitjournal{ApJL}

\shorttitle{Magnetic Switchback in the Solar Corona}
\shortauthors{Telloni et al.}

\begin{document}

\title{Observation of Magnetic Switchback in the Solar Corona}

\correspondingauthor{Daniele Telloni}
\email{daniele.telloni@inaf.it}

\author[0000-0002-6710-8142]{Daniele Telloni}
\altaffiliation{These authors contributed equally to this work.}
\affil{National Institute for Astrophysics, Astrophysical Observatory of Torino, Via Osservatorio 20, I-10025 Pino Torinese, Italy}
\author[0000-0002-4642-6192]{Gary P. Zank}
\altaffiliation{These authors contributed equally to this work.}
\affil{Center for Space Plasma and Aeronomic Research, University of Alabama in Huntsville, Huntsville, AL 35805, USA}
\affil{Department of Space Science, University of Alabama in Huntsville, Huntsville, AL 35805, USA}
\author[0000-0002-5365-7546]{Marco Stangalini}
\affil{Italian Space Agency, Via del Politecnico snc, I-00133 Roma, Italy}
\author[0000-0003-1759-4354]{Cooper Downs}
\affil{Predictive Science Inc., San Diego, CA 92121, USA}
\author[0000-0001-9581-4821]{Haoming Liang}
\affil{Center for Space Plasma and Aeronomic Research, University of Alabama in Huntsville, Huntsville, AL 35805, USA}
\author[0000-0002-7203-0730]{Masaru Nakanotani}
\affil{Center for Space Plasma and Aeronomic Research, University of Alabama in Huntsville, Huntsville, AL 35805, USA}
\author[0000-0003-1962-9741]{Vincenzo Andretta}
\affil{National Institute for Astrophysics, Astronomical Observatory of Capodimonte, Salita Moiariello 16, I-80131 Napoli, Italy}
\author[0000-0003-4155-6542]{Ester Antonucci}
\affil{National Institute for Astrophysics, Astrophysical Observatory of Torino, Via Osservatorio 20, I-10025 Pino Torinese, Italy}
\author[0000-0002-5981-7758]{Luca Sorriso-Valvo}
\affil{Swedish Institute of Space Physics, \AA ngstr\"om Laboratory, L\"agerhyddsv\"agen 1, SE-751 21 Uppsala, Sweden}
\affil{National Research Council, Institute for the Science and Technology of Plasmas, Via Amendola 122/D, I-70126 Bari, Italy}
\author[0000-0003-1549-5256]{Laxman Adhikari}
\affil{Center for Space Plasma and Aeronomic Research, University of Alabama in Huntsville, Huntsville, AL 35805, USA}
\author[0000-0002-4299-0490]{Lingling Zhao}
\affil{Center for Space Plasma and Aeronomic Research, University of Alabama in Huntsville, Huntsville, AL 35805, USA}
\author[0000-0002-6433-7767]{Raffaele Marino}
\affil{Laboratoire de M\'ecanique des Fluides et d'Acoustique, Centre National de la Recherche Scientifique, \'Ecole Centrale de Lyon, Universit\'e Claude Bernard Lyon 1, INSA de Lyon, F-69134 \'Ecully, France}
\author[0000-0002-1017-7163]{Roberto Susino}
\affil{National Institute for Astrophysics, Astrophysical Observatory of Torino, Via Osservatorio 20, I-10025 Pino Torinese, Italy}
\author[0000-0002-5467-6386]{Catia Grimani}
\affil{University of Urbino Carlo Bo, Department of Pure and Applied Sciences, Via Santa Chiara 27, I-61029 Urbino, Italy}
\affil{National Institute for Nuclear Physics, Section in Florence, Via Bruno Rossi 1, I-50019 Sesto Fiorentino, Italy}
\author[0000-0002-2464-1369]{Michele Fabi}
\affil{University of Urbino Carlo Bo, Department of Pure and Applied Sciences, Via Santa Chiara 27, I-61029 Urbino, Italy}
\affil{National Institute for Nuclear Physics, Section in Florence, Via Bruno Rossi 1, I-50019 Sesto Fiorentino, Italy}
\author[0000-0003-2647-117X]{Raffaella D'Amicis}
\affil{National Institute for Astrophysics, Institute for Space Astrophysics and Planetology, Via del Fosso del Cavaliere 100, I-00133 Roma, Italy}
\author[0000-0003-1059-4853]{Denise Perrone}
\affil{Italian Space Agency, Via del Politecnico snc, I-00133 Roma, Italy}
\author[0000-0002-2152-0115]{Roberto Bruno}
\affil{National Institute for Astrophysics, Institute for Space Astrophysics and Planetology, Via del Fosso del Cavaliere 100, I-00133 Roma, Italy}
\author[0000-0002-3559-5273]{Francesco Carbone}
\affil{National Research Council, Institute of Atmospheric Pollution Research, c/o University of Calabria, I-87036 Rende, Italy}
\author[0000-0002-9874-2234]{Salvatore Mancuso}
\affil{National Institute for Astrophysics, Astrophysical Observatory of Torino, Via Osservatorio 20, I-10025 Pino Torinese, Italy}
\author[0000-0001-9921-1198]{Marco Romoli}
\affil{University of Florence, Department of Physics and Astronomy, Via Giovanni Sansone 1, I-50019 Sesto Fiorentino, Italy}
\author[0000-0001-6273-8738]{Vania Da Deppo}
\affil{National Research Council, Institute for Photonics and Nanotechnologies, Via Trasea 7, I-35131 Padova, Italy}
\author[0000-0002-2789-816X]{Silvano Fineschi}
\affil{National Institute for Astrophysics, Astrophysical Observatory of Torino, Via Osservatorio 20, I-10025 Pino Torinese, Italy}
\author[0000-0002-5778-2600]{Petr Heinzel}
\affil{Czech Academy of Sciences, Astronomical Institute, Fri\v{c}ova 298, CZ-25165 Ond\v{r}ejov, Czechia}
\author[0000-0001-9670-2063]{John D. Moses}
\affil{National Aeronautics and Space Administration, Headquarters, Washington, DC 20546, USA}
\author[0000-0003-2007-3138]{Giampiero Naletto}
\affil{University of Padua, Department of Physics and Astronomy, Via Francesco Marzolo 8, I-35131 Padova, Italy}
\author[0000-0002-9459-3841]{Gianalfredo Nicolini}
\affil{National Institute for Astrophysics, Astrophysical Observatory of Torino, Via Osservatorio 20, I-10025 Pino Torinese, Italy}
\author[0000-0003-3517-8688]{Daniele Spadaro}
\affil{National Institute for Astrophysics, Astrophysical Observatory of Catania, Via Santa Sofia 78, I-95123 Catania, Italy}
\author[0000-0001-7298-2320]{Luca Teriaca}
\affil{Max Planck Institute for Solar System Research, Justus-von-Liebig-Weg 3, D-37077 G\"ottingen, Germany}
\author[0000-0001-9014-614X]{Federica Frassati}
\affil{National Institute for Astrophysics, Astrophysical Observatory of Torino, Via Osservatorio 20, I-10025 Pino Torinese, Italy}
\author[0000-0002-0764-7929]{Giovanna Jerse}
\affil{National Institute for Astrophysics, Astronomical Observatory of Trieste, Localit\`a Basovizza 302, I-34149 Trieste, Italy}
\author[0000-0001-8244-9749]{Federico Landini}
\affil{National Institute for Astrophysics, Astrophysical Observatory of Torino, Via Osservatorio 20, I-10025 Pino Torinese, Italy}
\author[0000-0002-3789-2482]{Maurizio Pancrazzi}
\affil{National Institute for Astrophysics, Astrophysical Observatory of Torino, Via Osservatorio 20, I-10025 Pino Torinese, Italy}
\author[0000-0002-2433-8706]{Giuliana Russano}
\affil{National Institute for Astrophysics, Astronomical Observatory of Capodimonte, Salita Moiariello 16, I-80131 Napoli, Italy}
\author[0000-0002-5163-5837]{Clementina Sasso}
\affil{National Institute for Astrophysics, Astronomical Observatory of Capodimonte, Salita Moiariello 16, I-80131 Napoli, Italy}
\author[0000-0003-4052-9462]{David Berghmans}
\affil{Solar-Terrestrial Centre of Excellence, Solar Influences Data Analysis Center, Royal Observatory of Belgium, Avenue Circulaire 3, B-1180 Brussels, Belgium}
\author[0000-0003-0972-7022]{Fr\'ed\'eric Auch\`ere}
\affil{Universit\'e Paris-Saclay, Centre National de la Recherche Scientifique, Institut d'Astrophysique Spatiale, B\^atiment 121, F-91405 Orsay, France}
\author[0000-0003-1294-1257]{Regina Aznar Cuadrado}
\affil{Max Planck Institute for Solar System Research, Justus-von-Liebig-Weg 3, D-37077 G\"ottingen, Germany}
\author[0000-0002-9270-6785]{Lakshmi P. Chitta}
\affil{Max Planck Institute for Solar System Research, Justus-von-Liebig-Weg 3, D-37077 G\"ottingen, Germany}
\author[0000-0001-9457-6200]{Louise Harra}
\affil{Physical Meteorological Observatory in Davos, World Radiation Center, Dorfstrasse 33, CH-7260 Davos Dorf, Switzerland}
\affil{Swiss Federal Institute of Technology in Z\"urich, H\"onggerberg Campus, Wolfgang-Pauli-Strasse 27, CH-8093 Z\"urich, Switzerland}
\author[0000-0002-2265-1803]{Emil Kraaikamp}
\affil{Solar-Terrestrial Centre of Excellence, Solar Influences Data Analysis Center, Royal Observatory of Belgium, Avenue Circulaire 3, B-1180 Brussels, Belgium}
\author[0000-0003-3137-0277]{David M. Long}
\affil{Mullard Space Science Laboratory, University College London, Holmbury St. Mary, RH5 6NT Dorking, UK}
\author[0000-0002-7762-5629]{Sudip Mandal}
\affil{Max Planck Institute for Solar System Research, Justus-von-Liebig-Weg 3, D-37077 G\"ottingen, Germany}
\author[0000-0003-1438-1310]{Susanna Parenti}
\affil{Universit\'e Paris-Saclay, Centre National de la Recherche Scientifique, Institut d'Astrophysique Spatiale, B\^atiment 121, F-91405 Orsay, France}
\author[0000-0002-0397-2214]{Gabriel Pelouze}
\affil{Universit\'e Paris-Saclay, Centre National de la Recherche Scientifique, Institut d'Astrophysique Spatiale, B\^atiment 121, F-91405 Orsay, France}
\author[0000-0001-9921-0937]{Hardi Peter}
\affil{Max Planck Institute for Solar System Research, Justus-von-Liebig-Weg 3, D-37077 G\"ottingen, Germany}
\author[0000-0002-6097-374X]{Luciano Rodriguez}
\affil{Solar-Terrestrial Centre of Excellence, Solar Influences Data Analysis Center, Royal Observatory of Belgium, Avenue Circulaire 3, B-1180 Brussels, Belgium}
\author[0000-0001-6060-9078]{Udo Sch\"uhle}
\affil{Max Planck Institute for Solar System Research, Justus-von-Liebig-Weg 3, D-37077 G\"ottingen, Germany}
\author[0000-0002-7669-5078]{Conrad Schwanitz}
\affil{Physical Meteorological Observatory in Davos, World Radiation Center, Dorfstrasse 33, CH-7260 Davos Dorf, Switzerland}
\affil{Swiss Federal Institute of Technology in Z\"urich, H\"onggerberg Campus, Wolfgang-Pauli-Strasse 27, CH-8093 Z\"urich, Switzerland}
\author[0000-0002-3281-4223]{Phil J. Smith}
\affil{Mullard Space Science Laboratory, University College London, Holmbury St. Mary, RH5 6NT Dorking, UK}
\author[0000-0002-5022-4534]{Cis Verbeeck}
\affil{Solar-Terrestrial Centre of Excellence, Solar Influences Data Analysis Center, Royal Observatory of Belgium, Avenue Circulaire 3, B-1180 Brussels, Belgium}
\author[0000-0002-2542-9810]{Andrei N. Zhukov}
\affil{Solar-Terrestrial Centre of Excellence, Solar Influences Data Analysis Center, Royal Observatory of Belgium, Avenue Circulaire 3, B-1180 Brussels, Belgium}
\affil{Skobeltsyn Institute of Nuclear Physics, Moscow State University, 119992 Moscow, Russia}

\begin{abstract}
Switchbacks are sudden, large radial deflections of the solar wind magnetic field, widely revealed in interplanetary space by the Parker Solar Probe. The switchbacks' formation mechanism and sources are still unresolved, although candidate mechanisms include Alfv\'enic turbulence, shear-driven Kelvin-Helmholtz instabilities, interchange reconnection, and geometrical effects related to the Parker spiral. This Letter presents observations from the Metis coronagraph onboard Solar Orbiter of a single large propagating S-shaped vortex, interpreted as first evidence of a switchback in the solar corona. It originated above an active region with the related loop system bounded by open-field regions to the East and West. Observations, modeling, and theory provide strong arguments in favor of the interchange reconnection origin of switchbacks. Metis measurements suggest that the initiation of the switchback may also be an indicator of the origin of slow solar wind.
\end{abstract}


\section{Introduction}
\label{sec:introduction}
The solar wind is a continuous flow of charged particles streaming from the Sun's outermost atmosphere, the solar corona, into interplanetary space \citep{1972cesw.book.....H}. It is characterized by the coexistence of large-scale structures of solar origin, turbulent fluctuations, magnetohydrodynamic (MHD) and kinetic plasma waves and instabilities, associated with physical processes such as magnetic reconnection, shocks, and a broad range of kinetic processes, which result in particle heating and energization \citep{2013LRSP...10....2B}.

The solar wind acceleration mechanisms, its complex dynamics and interaction with the solar magnetic field, and the observed plasma heating are still outstanding questions in heliophysics. One interesting solar wind feature emerging from in-situ spacecraft measurements is the puzzling existence of abrupt, temporary magnetic field reversals, named magnetic {\it switchbacks}.

First observed in the outer heliosphere by the Ulysses spacecraft in high-latitude fast solar wind \citep{1999GeoRL..26..631B,2004JGRA..109.3104Y}, Switchbacks are markedly Alfv\'enic, pressure-balanced structures, characterized by constant temperature and magnetic field magnitude, and associated with substantial acceleration of the plasma. Later analysis of Helios data in the ecliptic fast solar wind showed abundance of such structures in the inner heliosphere \citep{2018MNRAS.478.1980H}. More recently, extensive measurements by Parker Solar Probe \citep[PSP,][]{2016SSRv..204....7F} confirmed that the presence of switchbacks increases dramatically near the Sun \citep{2019Natur.576..237B,2019Natur.576..228K}, and allowed a thorough study of their characteristics.

Switchbacks have been described as three-dimensional, elongated S-shaped structures with a high aspect ratio in the radial direction \citep{2020ApJS..246...45H}. The typical deflection angle with respect to the heliospheric magnetic field is broadly distributed \citep{2020ApJS..246...39D}, and the duration of the reversals ranges between seconds and hours \citep{2020ApJS..246...39D,2021A&A...650A...1L}. At close distances from the Sun, switchbacks appear in dense, irregular clusters that alternate with quiet periods of stable field polarity and smaller magnetic fluctuations, with a modulation on a time scale of a few hours \citep{2021ApJ...923..174B}. The size and spatial distribution of the switchback patches, as well as the plasma characteristics within them, are compatible with the coronal magnetic structure determined by solar supergranulation \citep{2021ApJ...923..174B}. Switchbacks are associated with enhanced turbulence energy transfer \citep{2020ApJ...904L..30B,2020ApJ...905..142P,2021ApJ...922L..11H,2021ApJ...912...28M} and magnetic reconnection was observed at their edges \citep{2021A&A...650A...5F}. On the other hand, because of their limited size, energetic particle propagation along the interplanetary magnetic field does not seem to be affected by their presence \citep{2021A&A...650L...4B}.

It is however expected that switchbacks may play a fundamental role in driving processes that ultimately heat the solar wind. Despite the numerous and thorough experimental studies, the nature of the mechanisms generating the switchbacks is still being debated. For example, it is not clear if they are driven by processes in the lower solar atmosphere \citep{2021ApJ...911...75M} or self-consistently generated in the solar wind. While their overall occurrence characteristics observed by PSP so far better support a possible origin in the coronal transition region rather than in situ \citep{2021ApJ...919...60M,2021ApJ...923..174B,2021ApJ...919...96F}, several competing models have been proposed. For example, switchbacks could be the signature of frequent interchange reconnection events in the solar corona \citep{1999ESASP.446..675V,2009ApJ...691...61P,2020ApJ...894L...4F,2020ApJ...903....1Z,2020ApJ...896L..18S,2021A&A...650A...2D,2021ApJ...917..110L}. According to a different proposed mechanism of coronal generation, they could be associated with the motion of coronal magnetic field footpoints from the slow to the fast wind sectors \citep{2021ApJ...909...95S}. The observed coupling with solar supergranulation \citep{2021ApJ...923..174B} seems to suggest that switchbacks are nonlinear Alfv\'enic structures somehow associated with the global circulation of open magnetic flux at the solar surface \citep{2020ApJ...894L...4F,2021PhPl...28h0501Z}. On the other hand, MHD numerical simulations suggest that switchbacks may be Alfv\'enic structures originating in the lower corona and propagating out into the heliosphere \citep{2015ApJ...802...11M,2020ApJS..246...32T,2021A&A...647A..18J}, or may be related with velocity-shear driven dynamics \citep{2006GeoRL..3314101L,2020ApJ...902...94R}. From the same perspective, switchbacks may be generated self-consistently during the solar wind expansion of turbulent fluctuations \citep{2020ApJ...891L...2S,2021ApJ...915...52S}.

The variety of proposed models highlights the complex nature of switchbacks, and the limited vantage point provided by single-spacecraft, in-situ experimental observations available so far. Remote observation of switchbacks in the corona could provide a more complete description of their space-time dynamics, and help validate the models. However, detecting switchbacks in the extended corona with remote-sensing instruments is challenging and involves high spatial resolution coronagraphs. Such an endeavor is further complicated by the absence of coronal magnetic field measurements, so switchbacks can only be revealed by seeking their associated plasma counterparts.

This Letter reports observations from Metis \citep{2020A&A...642A..10A}, the coronagraph onboard Solar Orbiter \citep[SO,][]{2020A&A...642A...1M}, showing the first evidence of a magnetic switchback in the solar corona. A thorough data analysis, complemented by a stringent theoretical description and detailed modeling of the event and of the coronal magnetic field, strongly supports the interpretation of the switchback being generated by interchange reconnection between the coronal loops formed above an active region and the nearby open-field regions. This is also particularly relevant to the slow solar wind, whose origin, still widely debated, and seems likely to be related to interchange reconnection events continuously occurring at the boundaries between coronal holes and loop systems.

\section{Results}
\label{sec:results}
\subsection{Metis Observations}
\label{sec:results_metis_observations}
\begin{figure*}[h]
	\begin{center}
		\includegraphics[width=\linewidth]{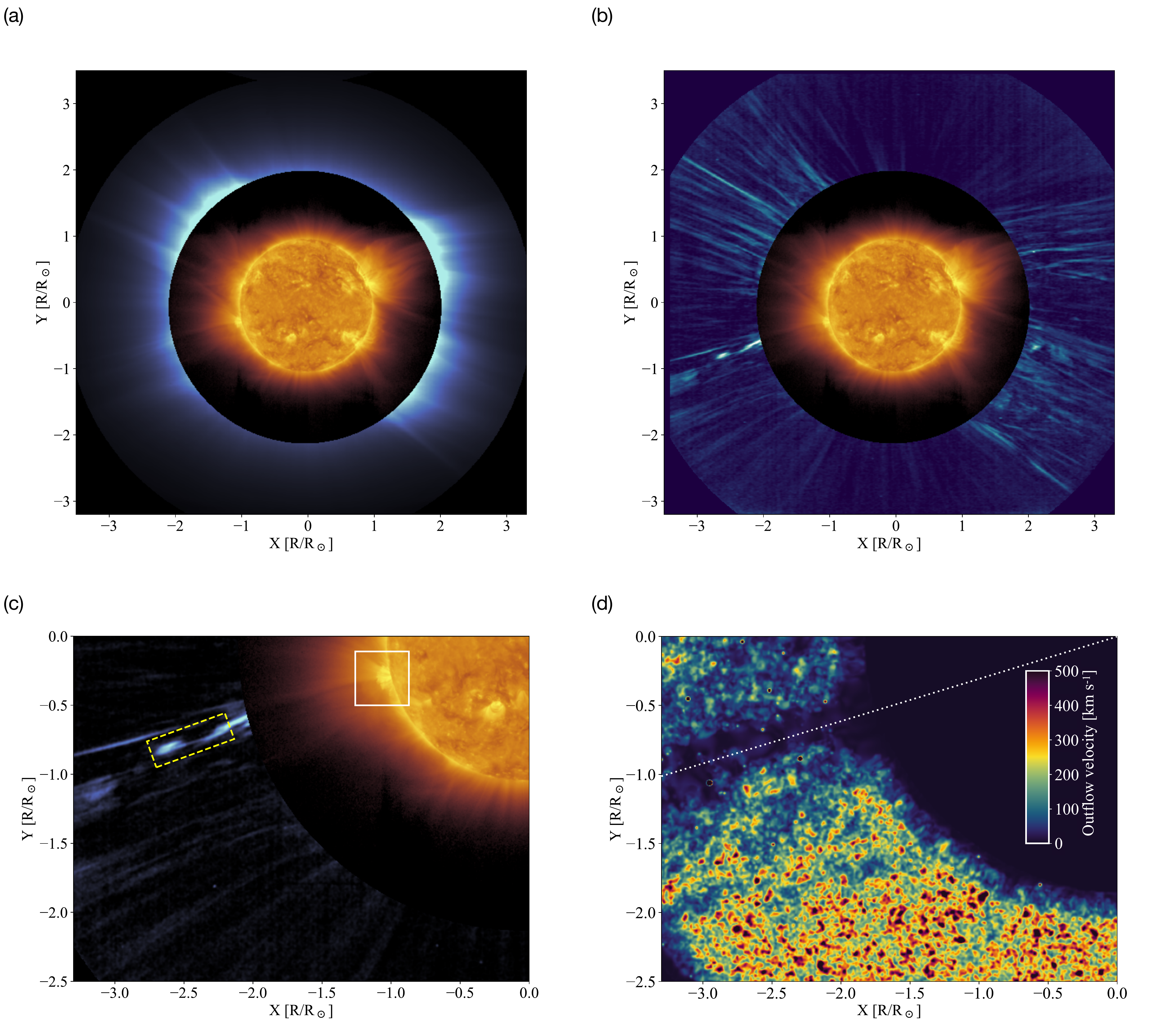}
	\end{center}
	\caption{(a) Composite of the Metis-observed tB image of the solar corona within $3.5$ R$_{\odot}$ FOV (blue) and the EUI image of the ultraviolet emission (yellow) as viewed from the SO vantage point on March $25$, $2022$ at $20$$:$$39$ UT. (b) Same as (a), but with the Metis observations radially filtered with the SiRGraF algorithm. (c) Zoom-in of (b); overlaid is the rectangular ROI where the time--distance analysis on the S-shaped structure was performed, and a white square indicating the loop system associated with AR $12972$. (d) Coronal velocity map in the same FOV as in (c); the white dotted line marks the latitude relative to the S-shaped feature.}
	\label{fig:metis_eui}
\end{figure*}

On March $25$, $2022$, during SO's first close approach to the Sun at a heliocentric distance of about $0.32$ au, the Metis coronagraph observed the solar corona in visible light ($580-640$ nm), in the annular field of view (FOV) from $1.9$ to $3.5$ R$_{\odot}$ (R$_{\odot}=696$ Mm). From $20$$:$$11$ to $20$$:$$52$ UT, Metis acquired $120$ $1024$$\times$$1024$-pixel coronal total brightness (tB) images, with spatial resolution of $4.7$ Mm ($=0.0068$ R$_{\odot}$) and time cadence of $20$ seconds. Figure \ref{fig:metis_eui}(a) displays a composite image of the solar corona on March $25$, $2022$ at $20$$:$$39$ UT, obtained by combining the Metis tB image (in blue) with observations from the Extreme Ultraviolet Imager \citep[EUI,][]{2020A&A...642A...8R} on SO, taken in ultraviolet light ($17.4$ nm, in yellow). Figure \ref{fig:metis_eui}(b) shows the same view of the corona but with the Metis image reprocessed by the Simple Radial Gradient Filter (SiRGraF) algorithm proposed by \citet{2022SoPh..297...27P} to bring out the short-scale dynamic coronal structures. A peculiar S-shaped kink in the stream of plasma flowing off the Sun is clearly visible against the darker coronal background, above the southeastern solar limb. It is worth noting that this coronal feature also emerges by applying independent and different methods, such as a wavelet-based image enhancement algorithm \citep{2013ApJ...767..138T} and the Proper Orthogonal Decomposition \citep[POD,][]{1981trtu.proc..215L}. A zoom of the S-shaped structure is provided in Figure \ref{fig:metis_eui}(c) and delimited by the rectangular region of interest (ROI). The structure formed above the complex loop system related to the Active Region (AR) $12972$ (marked with a white square in the figure). Finally, Figure \ref{fig:metis_eui}(d) shows the velocity of coronal flows in the same FOV as Figure \ref{fig:metis_eui}(c). The large-scale coronal flow pattern is derived by tracking outwardly moving density enhancements in the Metis white-light images, via the Fourier Local Correlation Tracking \citep[FLCT,][]{2008ASPC..383..373F} technique. Noteworthy is the slower plasma stream at latitudes (indicated by a white dotted line) where the S-structure occurs.

\begin{figure*}[h]
	\begin{center}
		\includegraphics[width=0.5\linewidth]{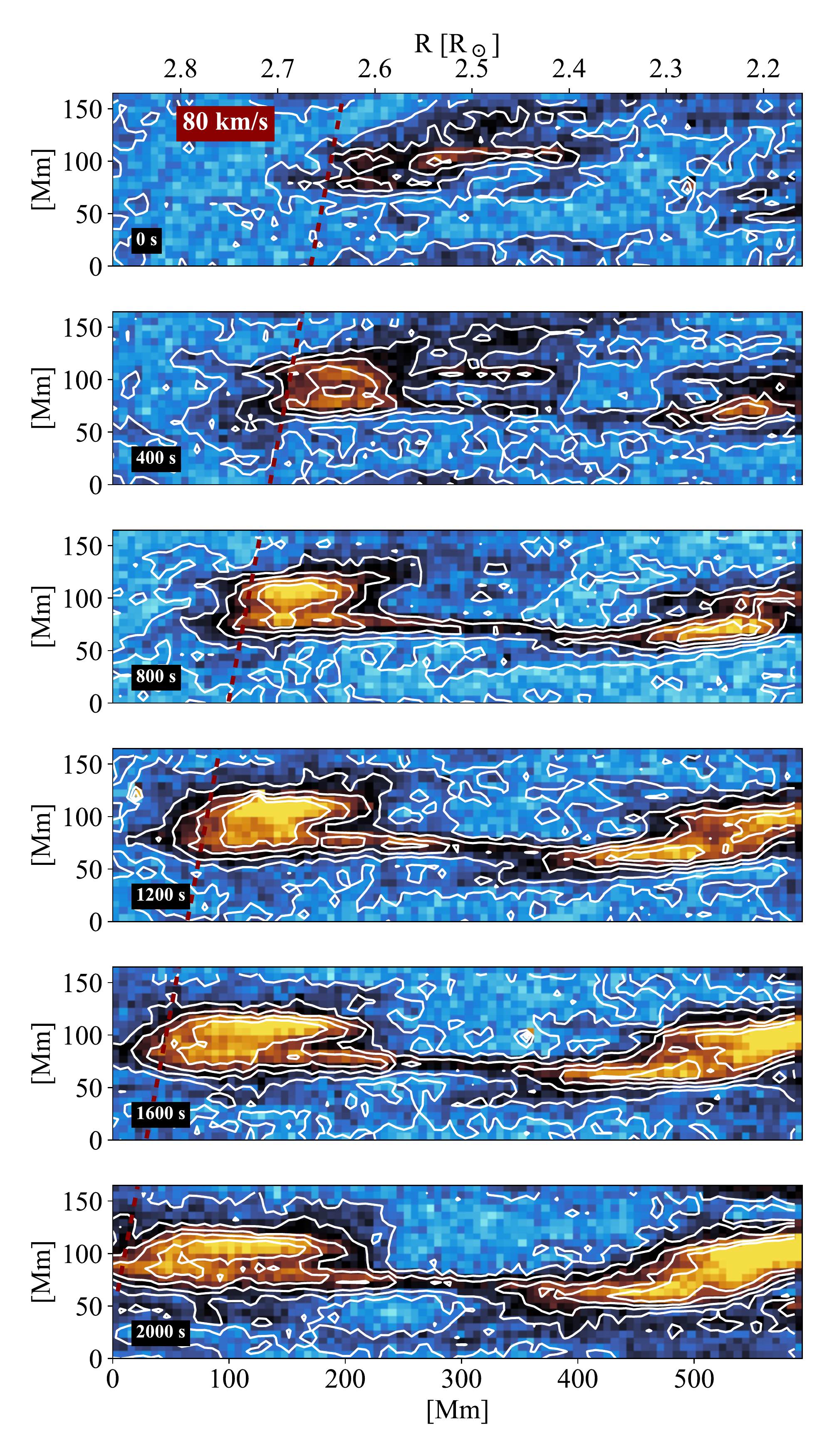}
	\end{center}
	\caption{Horizontally-oriented frames of the ROI overlaid in Figure \ref{fig:metis_eui} every $400$ seconds from $20$$:$$11$ UT on March $25$, $2022$. The upper y-axis indicates the distance from the Sun. The slope of the displayed red dashed line, connecting the radial positions of the main ripple front over time, indicates an almost constant propagation speed of $\sim80$ km s$^{-1}$.}
	\label{fig:roi}
\end{figure*}

The sequence of frames in the ROI, displayed in Figure \ref{fig:roi} every $400$ seconds from $20$$:$$11$ UT on March $25$, $2022$, shows the spatiotemporal evolution of the coronal feature under study. The direction along the structure is oriented horizontally and indicated with altitude above the Sun. The bright structure is observed initially well above the Metis occulted region, forming at a height of about $2.6$ R$_{\odot}$, and then propagates outward and folds back somewhat. At its full development, the structure acquires the peculiar ``S'' shape. The time--height plot allows an estimate of the propagation speed of the main ripple, which expands at a constant speed of about $80$ km s$^{-1}$ (red dashed line in Figure \ref{fig:roi}). This is, however, only a lower limit. Indeed, Metis can only estimate the structure's velocity component in the observed plane of the sky (POS), so that possible projection effects cannot be ruled out. In addition, the whole structure does not seem to propagate at the same rate, but rather to stretch slightly in the radial direction. This could be again due to projection on the POS of a very warped $3$D structure, or to actual velocity dispersion along the structure. Finally, the Metis images seem to suggest not only an outgoing but also an incoming flow (see, in particular, the frames at $800$ and $1200$ seconds after event initiation, where plasma filaments appear to protrude from the main ripple downward), although no firm conclusion can be drawn.

\begin{figure*}[h]
	\begin{center}
		\includegraphics[width=0.75\linewidth]{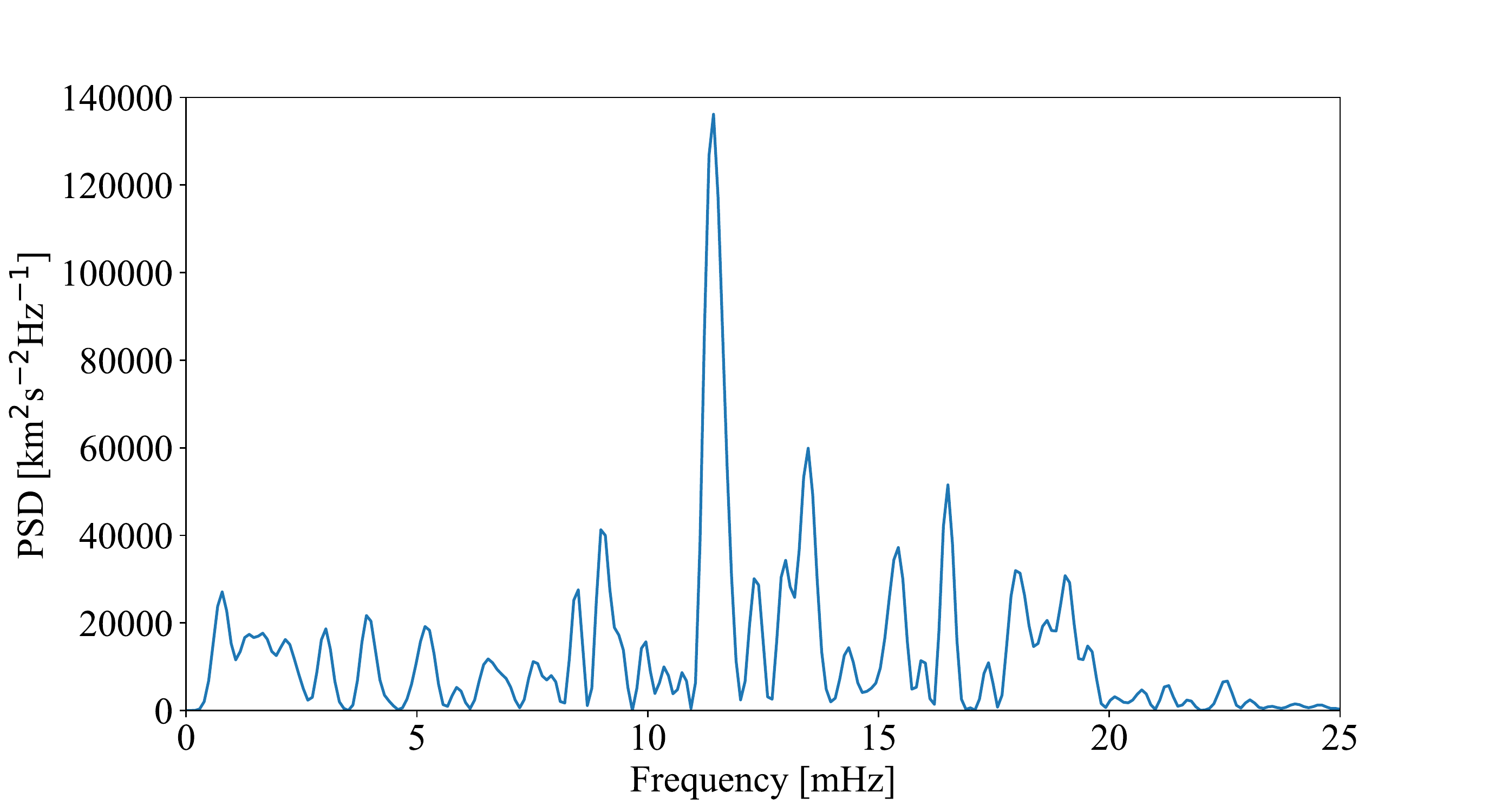}
	\end{center}
	\caption{Power spectral density of velocity fluctuations transverse to the main ripple propagation direction. A characteristic $12$ mHz frequency emerges clearly.}
	\label{fig:psd}
\end{figure*}

To investigate the vortical properties of the main ripple, which visual inspection seems to suggest, a spectral analysis was performed. Positioning in the rest reference frame of the evolving structure (i.e., at its barycenter) as it propagates outward, its transverse displacements with respect to the radial direction were evaluated and related to temporal fluctuations in velocity perpendicular to the structure. The corresponding Fourier power spectral density (Figure \ref{fig:psd}) shows clearly that there is a single dominant frequency (at $12$ mHz) with some sideband noise or less important fluctuations. Since $12$ mHz is not a common frequency, it is likely to be an intrinsic kink mode of the structure. It follows that the characteristic timescale of the vortex roll is about $80$ seconds. Hence, the ripple is essentially non-turbulent in nature, because otherwise a broadband spectrum should have been found.

\subsection{Coronal Magnetic Field Modeling}
\label{sec:results_coronal_magnetic_field_modeling}
The large-scale configuration of the coronal magnetic field during the observation period, and, most importantly, the local magnetic topology around the active region above which the S-shaped structure appears to emerge, is of crucial importance for context and for proper interpretation of the results obtained from the Metis data. A magnetic field reconstruction is accomplished by applying the $3$D MHD model developed by Predictive Science Incorporated (PSI) and based on the Wave-Turbulence-Driven (WTD) heating approach \citep{2018NatAs...2..913M}.

\begin{figure*}[h]
	\begin{center}
		\includegraphics[width=\linewidth]{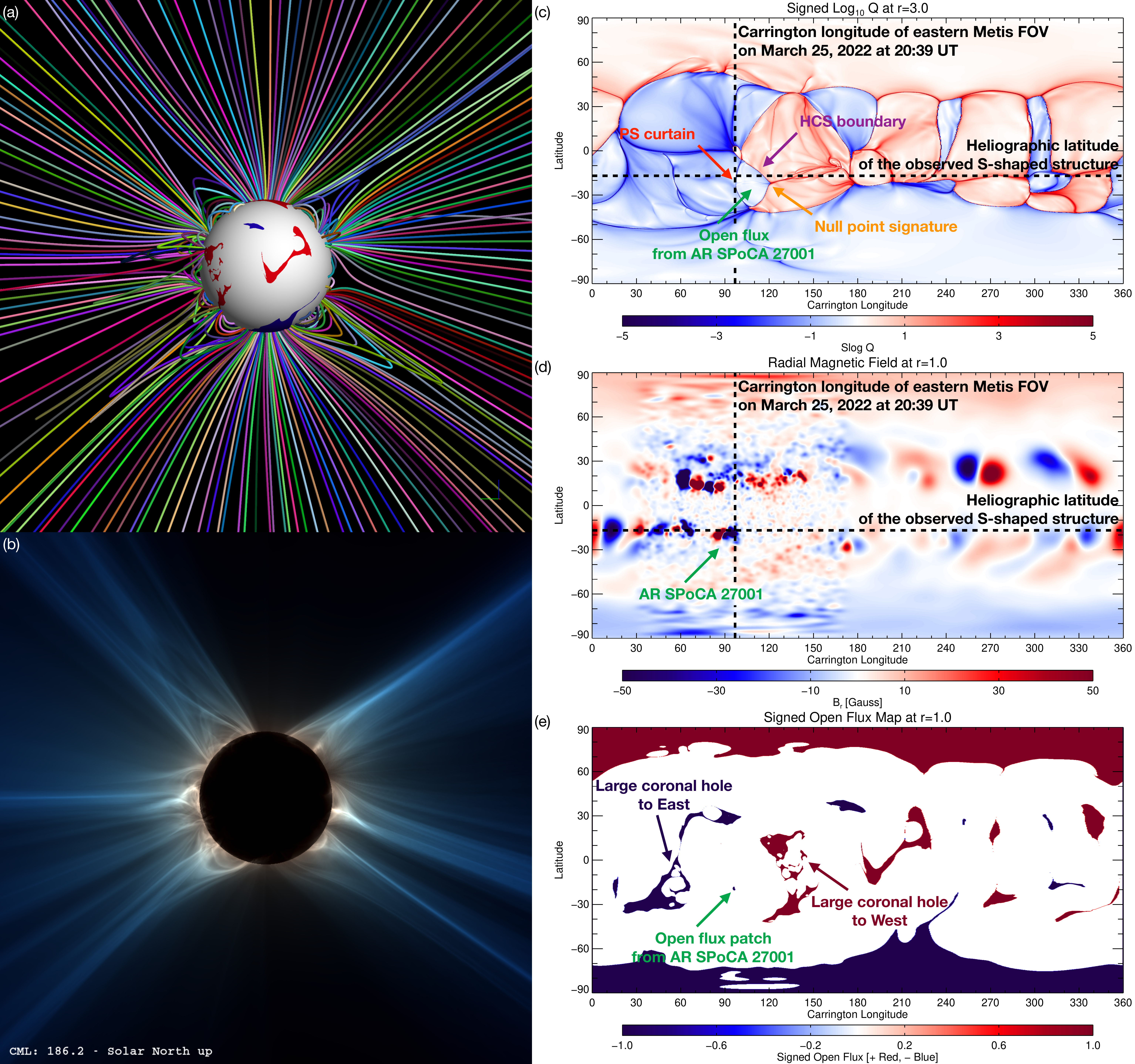}
	\end{center}
	\caption{Modeled magnetic field lines (a) and rendered squashing factor $Q$ (b) from the SO point of view on March $25$, $2022$. Carrington maps of $\log Q$ at $3$ R$_{\odot}$ (c), photospheric radial magnetic field (d), and open-field regions on the Sun's surface (e). The Carrington longitude of eastern Metis FOV on March $25$, $2022$ at $20$$:$$39$ UT and the heliographic latitude of the observed S-shaped structure are marked by black vertical and horizontal dashed lines, respectively. The AR $12972$ and the associated open flux are indicated by a green arrow. The latter maps to the PS curtain and the HCS (as indicated by color-coded arrows) in several places in latitude and longitude, forming a null point signature (denoted by the orange arrow).}
	\label{fig:psi}
\end{figure*}

The photospheric boundary conditions used for the extrapolation rely on Carrington maps of the measurements from the Helioseismic and Magnetic Imager \citep[HMI,][]{2012SoPh..275..207S} onboard the Solar Dynamics Observatory \citep[SDO,][]{2012SoPh..275....3P}. In the region of interest, observed at the East limb on March $25$, $2022$, the magnetic field map resolution is improved by including data segments of high-resolution HMI vector data. The modeled magnetic field lines traced on the Metis POS are displayed (along with the open flux regions on the solar surface) in Figure \ref{fig:psi}(a). The squashing factor $Q$ \citep{2011ApJ...731..111T}, which highlights the separatrix arcs of the $3$D coronal magnetic field, is imaged in Figure \ref{fig:psi}(b). The separatrix web, made up of the separatrix and quasi-separatrix layers surrounding the heliospheric current sheet (HCS), can be better visualized by building maps of $\log Q$ at $3$ R$_{\odot}$ (Figure \ref{fig:psi}(c)). In these layers, the photospheric dynamics stresses the separatrices inducing current sheets and reconnection processes with subsequent release of coronal plasma. The released plasma is proposed to contribute to the slow wind observed in the heliosphere in the zones around the current sheet \citep{2012SSRv..172..169A}. West of AR $12972$ (located exactly at the heliographic coordinates of the Metis-observed S-shaped feature), the separatrix between a tiny positive coronal hole, which is connected to the active region and touches the HCS, and a larger positive coronal hole has the signature of a null where they collide with the HCS. A pseudostreamer (PS) separatrix curtain lies on the East of the active region. Figures \ref{fig:psi}(d) and (e) finally show the synoptic maps of the photospheric radial magnetic field and of the footprints of the field lines associated with outward (red) and inward (blue) open flux at $3$ R$_{\odot}$. A small patch of open field lines can be identified at the active region.

What clearly emerges from the MHD extrapolation of the global and local magnetic field is the complex loop system above the active region and the large magnetic arcs extending up to $2-3$ R$_{\odot}$ in the FOV of Metis, just where the S-shaped, vortical structure forms. More interestingly, the active region appears to be surrounded, both East and West, by open-field regions injecting plasma along the coronal plasma sheet. Essentially, the large helmet streamer formed above AR $12972$ is broken up slightly by the fluxes/thermodynamics of the strong ARs. This is vital to the interpretation of the nature and source mechanism of the kink-like structure observed with Metis, as discussed below.

\section{Discussion}
\label{sec:discussion}
\subsection{Switchback Driven by Interchange Reconnection}
\label{sec:switchback_interchange_reconnection}
Based on Metis measurements, the previous section has provided observational evidence of the shape of a plasma structure in the solar corona, the height at which it was formed, the way it was generated, its spatiotemporal evolution, its vortical properties, the local magnetic field topology associated with an underlying active region, and the coupling of the structure with a slow coronal plasma flow. As will be argued in the following, all this evidence concurs in supporting the interpretation of the observed structure as a magnetic switchback generated in the solar corona through interchange reconnection.

The process of magnetic reconnection of a loop and an open field line, known as interchange reconnection, may occur at the boundaries between closed- and open-field regions at the streamer--coronal hole interface, and at the edges of active regions surrounded by unipolar regions or small coronal holes. The mechanism is depicted in the cartoon of Figure 3 in \citet{2020ApJ...903....1Z}. Interchange reconnection events can trigger the formation of magnetic field deflections, which propagate both outward and inward, whenever they are launched in the sub-Alfv\'enic flow \citep{2020ApJ...903....1Z}. The most extreme switchbacks take the characteristic S-shape and are detected in situ as magnetic field reversals. The launch of switchbacks could also be accompanied by coronal jets \citep{2020ApJ...896L..18S}. As a result of interchange reconnection, coronal plasma can be additionally released as blobs or plasmoids in the solar wind \citep{1997ApJ...484..472S,1998ApJ...498L.165W}. The switchback generated by such a mechanism (likely higher up in the corona, between large-scale loops and the adjacent open-field lines surrounding an active region) will hence be a single structure characterized by a perturbed magnetic field having a wavy shape. It will be associated with upward and potentially also downward flow of plasma that is different from the coronal background. In addition, such a switchback will consist of a single characteristic mode.

All these expected characteristics are actually depicted in the Metis images investigated here. First, the observed structure is S-shaped (Figure \ref{fig:metis_eui}(c)). If the direction of the flow is thought to be controlled by the strong coronal magnetic field, the white-light enhancement can be used as a way to infer the magnetic field direction. Thus, an S-shaped enhancement can be associated with an S-shaped magnetic field region. Second, the time--distance image plot of Figure \ref{fig:roi} indicates the observation of both switchback-related forward and inward flows, as predicted by \citet{2020ApJ...903....1Z}. Third, the observed coronal feature originates exactly above AR $12972$ (Figure \ref{fig:psi}(d)), whose loop system, extending well above the Metis occulter, is surrounded by open-field regions, as shown by magnetic field extrapolations (Figure \ref{fig:psi}(e)). Thus, it appears to be the ideal site for an interchange reconnection event to launch a switchback. Furthermore, the switchback-associated plasma is denser than the surrounding ambient corona (Figure \ref{fig:metis_eui}(c)), probably due to the compressions associated with the magnetic reconnection process and/or the entrapment of denser loop material. And, finally, the observation that the presumed switchback forms at high altitude in the solar corona \footnote{No propagating intensity disturbances in the inner corona were observed by EUI at that time.} is suggested by the possible location of the separatrix field lines at the null point (as well as by the squashing factor synoptic map displayed in Figure \ref{fig:psi}(c)). Fourth, observation of the single well-defined structure in Metis images rules out a shear-related Kelvin-Helmholtz (KH) type of driving \citep[as proposed by][]{2020ApJ...902...94R} and favors the interchange reconnection scenario. Indeed, a KH instability would not generate one large single propagating vortex, but rather multiple spatially distributed structures along the shear flow, many of which could be expected to be closely spaced or even interacting \citep[as at the CME flank observed by][and in striking contrast to Metis observations]{2011ApJ...729L...8F}. Fifth, the presence of a single dominant timescale (Figure \ref{fig:psd}) characterizing the observed ripple is consistent with a non-turbulent description of the switchback \citep[as in the model by][]{2020ApJ...903....1Z,2021ApJ...917..110L}. This is not compatible with the KH origin of switchbacks, or with the idea that switchbacks are a natural outcome of evolving turbulence in an expanding flow \citep[as advanced by][]{2020ApJ...891L...2S}. Indeed, in both cases, switchbacks would include a superposition of modes mutually interacting nonlinearly, thus generating a turbulent flow, without a characteristic dominant frequency. According to this analysis, the structure observed by Metis looks like a very clear switchback initiated by an interchange reconnection event occurring above an active region higher up in corona.

To conclude, adopting the perspective that interchange reconnection generates a switchback and then asking what the observational and theoretical support might be once that hypothesis is adopted, support for the existence of switchbacks in the solar corona generated by interchange reconnection can be argued from both observational and theoretical perspectives.

\subsection{Switchback Modeling}
\label{sec:discussion_switchback_modeling}
\begin{figure*}[h]
	\begin{center}
		\includegraphics[width=\linewidth]{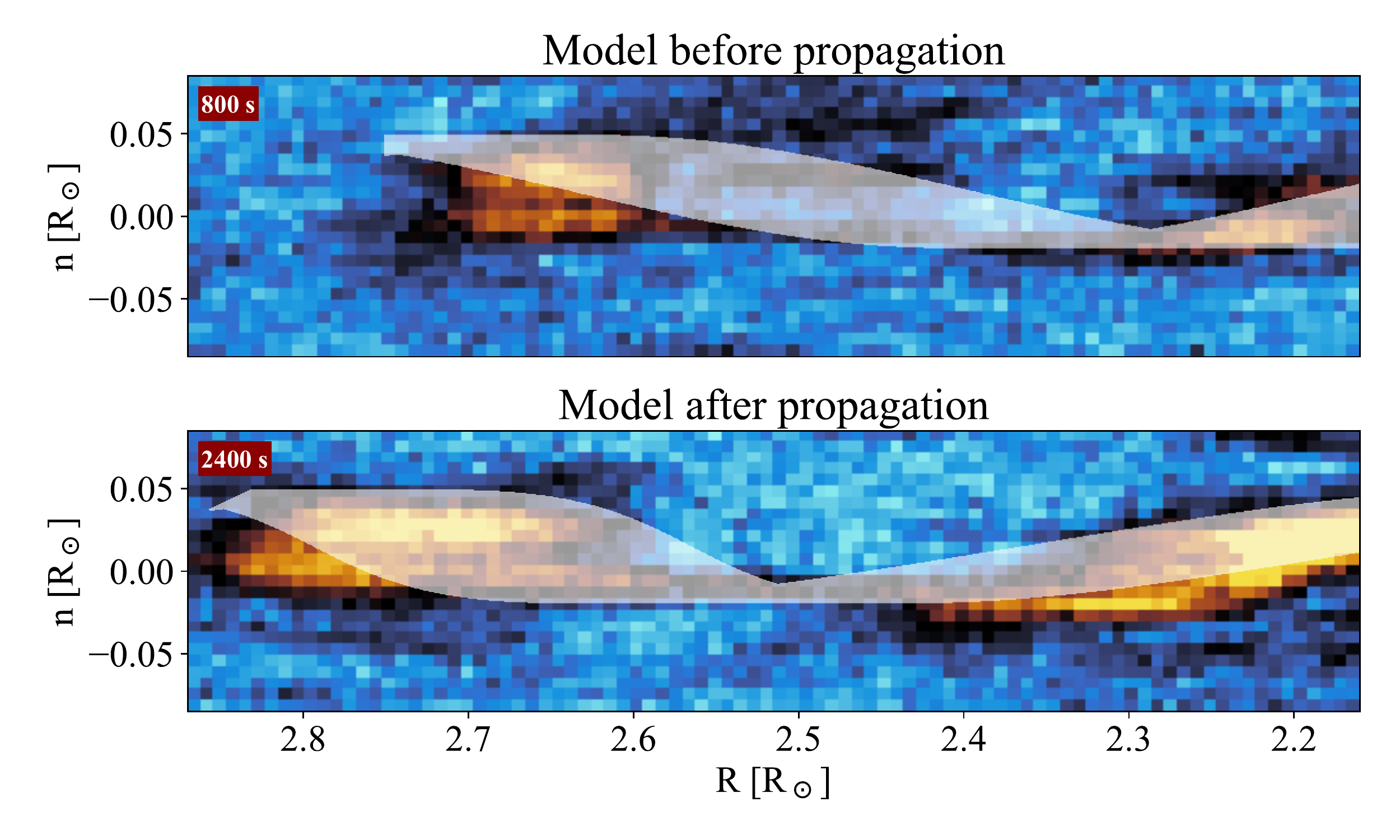}
	\end{center}
	\caption{MCMC-modeled switchback (white shape) before and after linear theory-based propagation, compared with Metis observations at $800$ and $2400$ seconds after initiation.}
	\label{fig:model}
\end{figure*}

To provide further clues to support the proposed interpretation and possibly to empirically validate the theoretical predictions by \citet{2020ApJ...903....1Z} against the Metis observations, the spatiotemporal evolution of the observed switchback structure was modeled using the Markov Chain Monte Carlo (MCMC) method. The shape of the structure was initially modeled with a sinusoidal function $n-n_{0}=A\sin\left[b\left(r-r_{0}\right)\right]$, where $n$ and $r$ are the positions in heliographic coordinates, $n_{0}$ and $r_{0}$ are the shifts in the $n$ and $r$ directions, respectively, $A$ is the amplitude of the perturbation in $n$, $b$ is the wave number of the sinusoidal function, and $d$ is a factor adjusting the width of the structure along the $r$ direction, i.e., $\big\vert r-r_{0}-\frac{1}{b}\arcsin\left(\frac{n-n_{0}}{A}\right)\big\vert<\frac{d}{2}$. In addition, to simulate the flow shear in the surrounding solar wind, a velocity gradient varying as $n$ ($U_{r}(n)=kn+U_{0}$, where $U_{r}(n)$ is the background solar wind speed, $U_{0}$ is a reference speed, and $k$ denotes the slope of the flow velocity) was also included. For simplicity, only the propagation in $r$ was considered. The initial structure was then allowed to evolve, based on the linear theory propagation function by \citet{2020ApJ...903....1Z}. The shape of the propagated structure was finally compared with Metis observations at $2400$ seconds after the switchback initiation, using the MCMC technique. The MCMC method is based on Bayesian inference \citep[e.g.,][]{2013sasd.book.....B} and was used to provide the most probable values of the free parameters $n_{0}$, $r_{0}$, $A$, $b$, $d$, and $k$ that lead to the optimal fit of the model to the observations \citep[this approach has been successfully applied in heliospheric studies by][where the interested reader can find more details on the application of the MCMC technique]{2019ApJ...886..144Z,2021ApJ...917..110L}. The results are displayed in Figure \ref{fig:model}, where the modeled switchback shape before and after propagation (white shape) is overlaid on the Metis image $800$ and $2400$ seconds after the switchback formation.

Although the model obviously cannot capture all features of the imaged switchback, the agreement with observations is striking. In particular, the modeled switchback elongates during propagation, which is consistent with the observations. Even more remarkable is that although the free parameters were set to best reproduce the shape of the $2400$-second switchback, the $800$-second structure so modeled also fits satisfactorily the observations. This is a strong indication of the validity of the linear theory proposed by \citet{2020ApJ...903....1Z} to describe the propagation of switchbacks in the corona, and greatly supports the scenario of having identified a magnetic switchback low in the solar corona for the first time.

\subsection{Switchbacks and the Slow Solar Wind}
\label{sec:discussion_switchback_slow_solar_wind}
The observation of switchbacks in the solar corona is closely related to the still unsolved problem of the origin of the slow solar wind. Indeed, one of the possible mechanisms for the origin of the slow coronal flows is that initially proposed by \citet{1999JGR...10419765F,2003JGRA..108.1157F} and later extended by \citet{2021PhPl...28h0501Z} by incorporating a mechanism for plasma heating. In this model \citep[well illustrated in Figure 1 of][]{2021PhPl...28h0501Z}, the large-scale loop plasma is heated to high temperatures via the dissipation of quasi-$2$D turbulence generated by the magnetic carpet. An interchange reconnection event with an open-field line at altitudes of $2-4$ R$_{\odot}$ liberates the already hot loop plasma onto an open-field coronal region, which then expands to reach supersonic and super-Alfv\'enic speeds. It appears thus evident that switchbacks and slow solar wind streams can be viewed as two manifestations of the same physical process, i.e., interchange reconnection. A number of clues supporting this possible scenario are present in the literature. For instance, the speed of the coronal blobs released by reconnection events \citep{1997ApJ...484..472S,1998ApJ...498L.165W} was found to be consistent with that of the slow wind \citep{2005A&A...435..699A}. Actually, quantifying the contribution of reconnection-related coronal blobs to the slow wind plasma has been the subject of several studies over the past decade \citep[e.g.,][]{2012SSRv..172..169A,2012SSRv..172..123W,2020ApJ...904..199W}. Furthermore, \citet{2008ApJ...676L.147H,2011ApJ...727L..13B,2011A&A...526A.137D,2016ApJ...823..145F,2017PASJ...69...47H} observed persistent coronal plasma upflows in association with quiescent active regions \citep[first identified by][]{1992PASJ...44L.155U} and interpreted this evidence as an indication of magnetic reconnection being the main driver of the slow wind. Furthermore, slow coronal flows have been observed to originate at the edges of coronal holes adjacent to closed-configuration equatorial streamers, and then stream along their flanks \citep{2005A&A...435..699A}. Finally, switchbacks are not observed in the middle of low-speed streams because, again in the interchange reconnection picture, these can occur only at the boundaries of slow and fast wind. Since both the loop-system and the open-field region consist of multiple magnetic field lines, interchange reconnection can occur in rapid succession, resulting in clustering of switchbacks \citep[consistent with PSP measurements, e.g.,][]{2020ApJS..246...39D} and in steady slow coronal flows. This process will continue while the loop system keeps emerging from the active region and until the magnetic footpoint motions ensure that the closed- and open-field regions are close enough to magnetically reconnect. Hence, the timing of the coronal switchback observed with Metis, as well as the height of its initiation, may be indicators of the origin of slow solar wind. As discussed in the previous section (Figures \ref{fig:metis_eui}(d) and \ref{fig:roi}), the switchback is observed to form at about $2.6$ R$_{\odot}$ and then expand at the low speed of $\sim80$ km s$^{-1}$ in a plasma stream slower than the adjacent ones. A remarkable relation of the switchback formation with the coronal region of slow plasma outflows has therefore been found, interestingly converging towards the interchange reconnection interpretation. Therefore, the study of switchbacks in the corona, especially where, how, and how often they are formed, seems to be very promising in disclosing the origin of the slow solar wind.

\acknowledgments
Solar Orbiter is a space mission of international collaboration between ESA and NASA, operated by ESA. D.T. was partially supported by the Italian Space Agency (ASI) under contract 2018-30-HH.0. G.P.Z., H.L., M.N., L.A., and L.-L.Z. acknowledge the partial support of a NASA Parker Solar Probe contract SV4-84017, an NSF EPSCoR RII-Track-1 Cooperative Agreement OIA-1655280, and a NASA IMAP grant through SUB000313/80GSFC19C0027. L.S.-V. was funded by the SNSA grants 86/20 and 145/18. L.P.C. gratefully acknowledges funding by the European Union. Views and opinions expressed are however those of the author(s) only and do not necessarily reflect those of the European Union or the European Research Council (grant agreement 101039844). Neither the European Union nor the granting authority can be held responsible for them. D.M.L. is supported by STFC Ernest Rutherford fellowship ST/R003246/1. S.P. acknowledges the funding by CNES through the MEDOC data and operations center. The Royal Observatory of Belgium team thanks the Belgian Federal Science Policy Office (BELSPO) for the provision of financial support in the framework of the PRODEX program of ESA under contract numbers 4000134474 and 4000136424. The Metis program is supported by ASI under contracts to the National Institute for Astrophysics and industrial partners. Metis was built with hardware contributions from Germany (Bundesministerium f\"ur Wirtschaft und Energie through the Deutsches Zentrum f\"ur Luft- und Raumfahrt e.V.), the Czech Republic (PRODEX) and ESA. The EUI instrument was built by CSL, IAS, MPS, MSSL/UCL, PMOD/WRC, ROB, LCF/IO with funding from the Belgian Federal Science Policy Office (BELSPO/PRODEX PEA 4000134088, 4000112292, 4000117262, and 4000134474); the Centre National d’\'Etudes Spatiales (CNES); the UK Space Agency (UKSA); the Bundesministerium f\"ur Wirtschaft und Energie (BMWi) through Deutsches Zentrum f\"ur Luft- und Raumfahrt (DLR); and the Swiss Space Office (SSO). The Metis data analyzed in this paper are available from the PI on request. ``EUI Data Release 5.0 2022-04'' is public and can be freely downloaded from the EUI website (\href{https://www.sidc.be/EUI/data/}{https://www.sidc.be/EUI/data/}) and from the Solar Orbiter Archive (\href{http://soar.esac.esa.int/soar/}{http://soar.esac.esa.int/soar/}). Inquiries regarding the coronal magnetic field extrapolation and the switchback modeling shown in the present study can be addressed to Cooper Downs (e-mail: \href{mailto:cdowns@predsci.com}{cdowns@predsci.com}) and Haoming Liang/Masaru Nakanotani (e-mails: \href{mailto:hl0045@uah.edu}{hl0045@uah.edu}; \href{mailto:mn0052@uah.edu}{mn0052@uah.edu}), respectively.
\par



\end{document}